\documentclass[]{svjour2}

\smartqed
\usepackage{amsmath,amssymb,graphicx}

\journalname{J Stat Phys}

\begin{document}

\title{Logarithmic Finite-Size Correction in Non-neutral Two-Component Plasma 
on Sphere}

\titlerunning{Non-neutral Two-Component Plasma on Sphere}

\author{Ladislav \v{S}amaj}

\institute{Institute of Physics, Slovak Academy of Sciences, 
D\'ubravsk\'a cesta 9, SK-84511 Bratislava, Slovakia \\
\email{Ladislav.Samaj@savba.sk}}

\date{Received:  / Accepted: }

\maketitle

\begin{abstract}
We consider a general two-component plasma of classical pointlike charges 
$+e$ ($e$ is say the elementary charge) and $-Z e$ (valency $Z=1,2,\ldots$), 
living on the surface of a sphere of radius $R$.
The system is in thermal equilibrium at the inverse temperature $\beta$, in
the stability region against collapse of oppositely charged particle pairs
$\beta e^2 < 2/Z$.
We study the effect of the system excess charge $Q e$ on the finite-size
expansion of the (dimensionless) grand potential $\beta\Omega$.
By combining the stereographic projection of the sphere onto an infinite 
plane, the linear response theory and the planar results for the second
moments of the species density correlation functions we show that for any  
$\beta e^2 < 2/Z$ the large-$R$ expansion of the grand potential is of the form 
$\beta\Omega \sim A_V R^2 + [\chi/6 - \beta (Qe)^2/2] \ln R$, 
where $A_V$ is the non-universal coefficient of the volume (bulk) part and 
the Euler number of the sphere $\chi=2$.
The same formula, containing also a non-universal surface term proportional
to $R$, was obtained previously for the disc domain ($\chi=1$), in the case of 
the symmetric $(Z=1)$ two-component plasma at the collapse point $\beta e^2=2$
and the jellium model $(Z\to 0)$ of identical $e$-charges in a fixed 
neutralizing background charge density at any coupling $\beta e^2$ 
being an even integer. 
Our result thus indicates that the prefactor to the logarithmic finite-size
expansion does not depend on the composition of the Coulomb fluid and 
its non-universal part $-\beta (Qe)^2/2$ is independent of the geometry of 
the confining domain. 


\end{abstract}

\renewcommand{\theequation}{1.\arabic{equation}}
\setcounter{equation}{0}

\section{Introduction} \label{Sect.Introduction}
Let a system of particles with {\em short-ranged} interactions in thermal 
equilibrium at the inverse temperature $\beta=1/(k_{\rm B}T)$, confined to 
a large two-dimensional (2D) domain of characteristic size $R$, be in its 
critical point.
According to the principle of conformal invariance 
\cite{Affleck86,Blote86,Cardy88}, the (dimensionless) grand potential
$\beta\Omega$ has a large-$R$ expansion
\begin{equation} \label{univ}
\beta\Omega = A_V R^2 + A_S R + B \ln R + \cdots ,
\end{equation}
where the volume and surface coefficients $A_V$ and $A_S$ are non-universal,
while the coefficient of the logarithmic term 
\begin{equation} \label{B1}
B = - \frac{c\chi}{6} 
\end{equation}
is universal, dependent only on the conformal anomaly number $c$ of 
the critical theory and the Euler number $\chi$ of the confining domain 
($\chi=1$ for a disk and $\chi=2$ for the surface of a sphere).

In this paper, we are concerned with 2D Coulomb fluids of classical
(i.e., non-quantum) pointlike charges interacting pairwisely by
the {\em long-ranged} logarithmic potential.
Two kinds of Coulomb models are of special interest.
The one-component plasma (OCP), or the jellium, consists of identical 
(say elementary) charges $e$ moving in a fixed neutralizing background charge 
density.
The symmetric two-component plasma (TCP) is a system of oppositely charged 
species $\pm e$.
In 2D, the thermodynamics and the particle correlation functions of both 
the OCP and the symmetric TCP depend on the only coupling constant 
$\Gamma=\beta e^2$.
The weak-coupling limit $\Gamma\to 0$ is treated exactly within the nonlinear 
Poisson-Boltzmann or linear Debye-H\"uckel mean-field theories \cite{Attard02}.
The 2D OCP is exactly solvable at $\Gamma=2$ by mapping onto free 
fermions \cite{Alastuey81,Jancovici81}.
The symmetric 2D TCP is also exactly solvable at the coupling $\Gamma=2$,
which corresponds to the collapse border for positive-negative pairs of 
pointlike charges, by mapping onto the free-fermion point of 
the Thirring model \cite{Cornu87,Gaudin85}.
For a review about exact results for 2D Coulomb systems, see 
Refs. \cite{Forrester98,Jancovici92}.

In the conducting regime, the long-range tail of the Coulomb potential 
induces screening and the electrical-field correlations become long-ranged 
\cite{Alastuey84,Jancovici95,Lebowitz84}.
As a consequence, the grand potential (or the free energy) of any Coulomb
system exhibits the universal logarithmic finite-size term of type 
(\ref{univ}).
For both the OCP and the symmetric TCP, the checks of the universal expansion 
were done in the weak coupling limit $\Gamma\to 0$ and at $\Gamma=2$,
for periodic boundary conditions \cite{Forrester91}, plain hard walls
\cite{Jancovici94}, ideal-conductor \cite{Jancovici96} and ideal-dielectric 
\cite{Jancovici01,Tellez01} boundaries, with the result
\begin{equation}
c = -1 .
\end{equation}
This $c$ is related to the Gaussian one \cite{Cardy90} by a change of sign.

A special case of the confining domain for the Coulomb system is
the surface of a sphere \cite{Caillol81,DiFrancesco94,Salazar16}.
For such geometry, by combining a stereographic projection of the sphere 
onto an infinite plane with linear response theory 
(TCP, Ref. \cite{Jancovici00a}) or with density functional method
(OCP, Ref. \cite{Jancovici00c}), the prefactor to the universal logarithmic
finite-size term was related to the second moment of the short-range part 
of the planar direct correlation function. 
Based on a renormalized Mayer expansion \cite{Deutsch74,Friedman62},
this quantity is known for both symmetric TCP \cite{Jancovici00b} and 
OCP \cite{Kalinay00}.
The case of an asymmetric TCP on a sphere was treated in Ref. \cite{Samaj01}.
All results mentioned up to now were derived for strictly neutral Coulomb 
systems. 

Recently, the symmetric 2D TCP of $\pm e$ charges, confined to a disk of
radius R and with a hard-core impurity of charge $Q e$ fixed at the disk
origin, was solved exactly at the collapse $\beta e^2=2$ point \cite{Ferrero14}.
The fixed impurity charge is screened on microscopic scale by counterions
from TCP, so the rest system can be considered as a non-neutral entity of
charge $Q e$.
It was shown that the grand potential still exhibits the finite-size expansion
of type (\ref{univ}) where the prefactor to the logarithmic term contains also 
the $Q$-dependent term:
\begin{equation} \label{B2}
B = \frac{1}{6} - Q^2 , \qquad (\chi = 1, \beta e^2=2) .
\end{equation}
This result is related to the minimal free-boson conformal field theory,
which is formally equivalent to the 2D TCP, formulated on the disk
\cite{Dotsenko04,Ginsparg90}.
Deforming the free-boson conformal theory by spreading out at infinity
a charge $Q e$, the prefactor to the logarithmic term was found, at an 
arbitrary coupling, in the form 
\begin{equation} \label{B3}
B = \frac{1}{6} - \frac{1}{2} \beta (Q e)^2 , \qquad 
(\chi = 1, \mbox{arbitrary $\beta e^2 < 2$}) .
\end{equation}
Note that the result (\ref{B2}) for the 2D TCP \cite{Ferrero14} is the special 
$\beta e^2=2$ case of this formula.
The coefficient $B$ is no longer universal, it depends on the inverse
temperature $\beta$ and the square of the excess charge $(Q e)^2$.

The case of the non-neutral 2D OCP confined to the disk was studied in 
Ref. \cite{Samaj17}.
For any coupling constant being an even integer, the mapping
of the system with an excess charge $Q e$ onto an anticommuting field theory 
formulated on a discrete chain provides for the free energy the large-$R$ 
expansion of type (\ref{univ}), with the coefficient to the logarithmic term 
$B$ exactly the same as in the relation (\ref{B3}).
This result indicates that the $B$-coefficient does not depend on the 
composition of the Coulomb system.

The finite-size expansions for non-neutral Coulomb fluids obtained till now
were restricted to the disk domain.
To investigate the effect of domain's geometry on the crucial $B$-coefficient,
we study in this work a non-neutral 2D Coulomb system living on the surface of 
a sphere of radius $R$.
In order to test also the independence of the coefficient $B$ on 
the charged species composition, we consider a general TCP of charges $+e$ 
and $-Z e$ (valency $Z=1,2,\ldots$) which involves as special cases
the symmetric TCP $(Z=1)$ as well as the OCP $(Z\to 0)$, after subtracting
the kinetic energy of species with charge $-Z e$.
By combining stereographic projection of the sphere onto a plane with linear 
response theory and using specific planar results for the second moments of 
the species density correlation functions of the asymmetric TCP derived in 
\cite{Samaj01}, it is shown that for the general TCP with an excess charge 
$Q e$ the $B$-coefficient takes the form
\begin{equation} \label{B4}
B = \frac{1}{3} - \frac{1}{2} \beta (Q e)^2 , \qquad 
(\chi = 2, \mbox{arbitrary $\beta e^2 < 2/Z$}) .
\end{equation}
This result supports the previous suggestion that the prefactor 
to the logarithmic finite-size term does not depend on the charge composition 
of the Coulomb system (in our case, the parameter $Z$).
Moreover, while the first universal term depends only on the shape of 
the confining domain, the non-universal part, depending on the inverse 
temperature $\beta$ and the square of the excess charge $(Q e)^2$, 
is the same for both disk and sphere geometries and therefore it presumably 
does not depend on domain's geometry. 

The paper is organized as follows.
The definition and basic relations for the general TCP living on the 
surface of a sphere is the subject of Sect. \ref{Sect.Sphere}.
Sect. \ref{Sect.Projection} reviews the stereographic projection of 
the system onto the one on an infinite surface. 
In Sect. \ref{Sect.Response}, the combination of linear-response arguments 
with the planar results for the second moments of the species density 
correlation functions \cite{Samaj01} leads to our main result (\ref{B4}).
A short recapitulation and concluding remarks about a phenomenological
explanation and generalization of the obtained results are given in 
Sect. \ref{Sect.Conclusion}.

\renewcommand{\theequation}{2.\arabic{equation}}
\setcounter{equation}{0}

\section{General TCP on a sphere} \label{Sect.Sphere}
Let $(\theta,\varphi)$ be the spherical coordinates of points on the surface 
of the sphere of radius $R$.
In Gauss units and with the vacuum dielectric constant $\varepsilon=1$,
the Coulomb potential $V(\theta)$ generated by a unit charge fixed 
at the north pole $\theta=0$ is given by \cite{Caillol81,DiFrancesco94} 
\begin{equation} \label{spherepot}
V(\theta) = - \ln\left[ \frac{2R}{L} \sin\left( \frac{\theta}{2} \right) 
\right] ,
\end{equation}
where $L$ is a length scale, $2R\sin(\theta/2)$ is the Euclidean distance from 
the north pole to the point $(\theta,\varphi)$.
In the limit $R\to\infty$, this potential reduces to the standard 2D
logarithmic one.
Two particles $i=1,2$ with charges $q_i$ and at spatial positions 
$(\theta_i,\varphi_i)$ interact by the potential
\begin{equation}
V_{12}(\tau) = - q_1 q_2 \ln\left[ \frac{2R}{L} 
\sin\left( \frac{\tau_{12}}{2} \right) \right] ,
\end{equation}
where $\tau_{12}$ is the angular distance between the points 1 and 2.
Using the vector representation 
${\bf r}_i=R(\sin\theta_i\cos\varphi_i,\sin\theta_i\sin\varphi_i,\cos\theta_i)$
$(i=1,2)$ and the scalar product formula 
${\bf r}_1\cdot {\bf r}_2 = R^2 \cos\tau_{12}$, one gets
\begin{equation} \label{angular}
\sin^2\left( \frac{\tau_{12}}{2} \right) = \frac{1}{2}
\left[ 1 - \sin\theta_1 \sin\theta_2 \cos(\varphi_1-\varphi_2) 
- \cos\theta_1 \cos\theta_2 \right] .
\end{equation}

The non-neutral model under consideration is the general TCP of positive 
$+e$ and negative $-Z e$ (valency $Z=1,2,\ldots$) charges, with an excess 
charge $Q e$.
Denoting the surface element of the sphere by 
\begin{equation} \label{element}
{\rm d}\sigma = R^2 {\rm d}(\cos\theta) {\rm d}\varphi ,
\end{equation}
the grand partition function is given by
\begin{subequations} \label{Xi}
\begin{equation} 
\Xi(\lambda_+,\lambda_-,R) = \sum_{N=0}^{\infty} 
\frac{\lambda_+^{Q+NZ}}{(Q+NZ)!} \frac{\lambda_-^N}{N!}
\int \prod_{i=1}^{Q+NZ} \frac{{\rm d}\sigma_i^+}{\ell^2}
\prod_{i=1}^N \frac{{\rm d}\sigma_i^-}{\ell^2} W_{Q+NZ,Z} 
\end{equation}
where $\lambda_+ = \exp(\beta\mu_+)$ and $\lambda_- = \exp(\beta\mu_+)$
are respectively the fugacities of $+e$ and $-Ze$ charges, $\ell$ stands for
the thermal de Broglie wavelength and
\begin{equation}
W_{Q+NZ,Z} = \frac{\prod_{(i<j)=1}^{Q+NZ} \left[ \frac{2R}{L} \sin\left( 
\frac{\tau_{ij}^{++}}{2}\right) \right]^{\Gamma}
\prod_{(i<j)=1}^N \left[ \frac{2R}{L} \sin\left( 
\frac{\tau_{ij}^{--}}{2}\right) \right]^{\Gamma Z^2}
}{\prod_{i=1}^{Q+NZ} \prod_{j=1}^N \left[ \frac{2R}{L} 
\sin\left( \frac{\tau_{ij}^{+-}}{2}\right) \right]^{\Gamma Z}}
\end{equation}
\end{subequations}
is the interaction Boltzmann factor of $Q+NZ$ particles with
charge $+e$ and $N$ particles with charge $-Ze$. 
Here, the dimensionless parameter $\Gamma\equiv \beta e^2$; for the symmetric 
$Z=1$ two-component plasma, it is equivalent to the coupling constant.
The (dimensionless) grand potential is defined as
\begin{equation} \label{grandpot}
\beta \Omega = - \ln\Xi .
\end{equation}
The 2D integrals in the expansion (\ref{Xi}) are stable (i.e., non-diverging)
against the collapse of oppositely charged particles if $\Gamma Z<2$.
To simplify the notation, we set the irrelevant lengths $L=\ell=1$.

The grand partition sum (\ref{Xi}) is the generating function for the total 
numbers $N_+^{(s)}$ of $+e$ charges and $N_-^{(s)}$ of $-Ze$ charges 
on the sphere according to
\begin{equation} \label{N}
N_+^{(s)} = \lambda_+ \frac{\partial}{\partial\lambda_+} \ln \Xi , \qquad
N_-^{(s)} = \lambda_- \frac{\partial}{\partial\lambda_-} \ln \Xi .
\end{equation}
The presence of the excess charge $Q e$ in the system is equivalent to the 
constraint
\begin{equation} \label{constraint}
e N_+^{(s)} - Z e N_-^{(s)} = Q e .  
\end{equation}
Because of the sphere symmetry all surface points are equivalent and
therefore the number densities of the species per unit surface are constant,
\begin{equation} \label{n}
n_+^{(s)} = \frac{N_+^{(s)}}{4\pi R^2} , \qquad 
n_-^{(s)} = \frac{N_-^{(s)}}{4\pi R^2} .
\end{equation}
In terms of the surface number densities, the condition (\ref{constraint})
is written as 
\begin{equation} \label{excess}
n_+^{(s)} - Z n_-^{(s)} = \frac{Q}{4\pi R^2} .  
\end{equation}

Let us consider the $N$th term in the expansion (\ref{Xi}).
Extracting all $R$-dependent parts, this term equals to
\begin{equation}
\lambda_+^Q R^{\Gamma Q^2/2} R^{Q(4-\Gamma)/2}
\left[ \lambda_+^Z \lambda_- R^{(1+Z)(4-\Gamma Z)/2} \right]^N  
\end{equation}
times a dimensionless $2N$-dimensional integral which depends on $\Gamma$
and $Z$.
Thus, 
\begin{equation} \label{scale}
\ln\Xi = Q \ln\lambda_+ + \left[ \frac{Q}{2}\left( 4-\Gamma \right)
+ \frac{\Gamma Q^2}{2} \right] \ln R + g(x) , 
\end{equation}
where the unknown function $g$ depends on $\lambda_+$, $\lambda_+$
and $R$ through the combination
\begin{equation}
x = \lambda_+^Z \lambda_- R^{(1+Z)(4-\Gamma Z)/2} .
\end{equation}

Based on the homogeneity relation (\ref{scale}), we get 
the following equalities
\begin{subequations}
\begin{eqnarray}
\lambda_+ \frac{\partial}{\partial\lambda_+} \ln\Xi & = & Q + Z x g'(x) , \\
\lambda_- \frac{\partial}{\partial\lambda_-} \ln\Xi & = & x g'(x) , \\
2 R \frac{\partial}{\partial R} \ln\Xi & = & (4-\Gamma) Q + \Gamma Q^2
+ (1+Z)(4-\Gamma Z) x g'(x) .
\end{eqnarray}
\end{subequations}
With regard to the definition of species number densities (\ref{N}) and
(\ref{n}), these relations imply
\begin{subequations} \label{rov}
\begin{eqnarray} 
n_+^{(s)} & = & \frac{Q}{4\pi R^2} + \frac{1}{2\pi R^2} 
\frac{Z}{(1+Z)(4-\Gamma Z)} \nonumber \\ & & \times
\left\{ R \frac{\partial}{\partial R} \ln\Xi 
- \frac{1}{2} \left[ (4-\Gamma)Q + \Gamma Q^2 \right] \right\} , \\
n_-^{(s)} & = & \frac{1}{2\pi R^2} \frac{1}{(1+Z)(4-\Gamma Z)} 
\nonumber \\ & & \times
\left\{ R \frac{\partial}{\partial R} \ln\Xi - \frac{1}{2} 
\left[ (4-\Gamma)Q + \Gamma Q^2 \right] \right\} .
\end{eqnarray} 
\end{subequations}
Note that the charge constraint (\ref{excess}) is automatically satisfied.

In the large-$R$ limit, $\ln\Xi$ behaves as
\begin{equation} \label{finite}
\ln \Xi = \beta P (4\pi R^2) + f(R) ,
\end{equation}
where $P$ is the bulk pressure of an infinite planar system and $f(R)$
a finite-size correction.
Since the sphere has no boundary, there is no term proportional to $R$
and therefore $f(R) = o(R)$.
Substituting (\ref{finite}) into (\ref{rov}) and taking the $R\to\infty$ 
limit, we obtain
\begin{subequations}
\begin{eqnarray}
n_+ = \frac{Z}{1+Z} \left( 1 - \frac{\Gamma Z}{4} \right)^{-1} \beta P , \\
n_- = \frac{1}{1+Z} \left( 1 - \frac{\Gamma Z}{4} \right)^{-1} \beta P ,
\end{eqnarray}
\end{subequations}
where $n_+$ and $n_-$ are the species densities of an infinite system
which satisfy the obvious neutrality condition
\begin{equation}
e n_+ - Z e n_- = 0 .
\end{equation}
The equation of state reads as
\begin{equation}
\beta P = \left( 1 - \frac{\Gamma Z}{4} \right) n
\end{equation}
with $n = n_+ + n_-$ being the total number density of charged particles. 
Finally, inserting the expansion (\ref{finite}) into (\ref{rov}) results in 
the couple of equations for the deviations of species densities on the sphere 
of radius $R$ from their asymptotic $R\to\infty$ planar values:
\begin{subequations} \label{rovnice}
\begin{eqnarray} 
n_+^{(s)} - n_+ & = & \frac{Q}{4\pi R^2} + \frac{1}{2\pi R^2} 
\frac{Z}{(1+Z)(4-\Gamma Z)} \nonumber \\ & & \times
\left\{ R \frac{\partial}{\partial R} f(R) 
- \frac{1}{2} \left[ (4-\Gamma)Q + \Gamma Q^2 \right] \right\} , \\
n_-^{(s)} - n_- & = & \frac{1}{2\pi R^2} \frac{1}{(1+Z)(4-\Gamma Z)}
\nonumber \\ & & \times 
\left\{ R \frac{\partial}{\partial R} f(R) - \frac{1}{2} 
\left[ (4-\Gamma)Q + \Gamma Q^2 \right] \right\} .
\end{eqnarray} 
\end{subequations}

\renewcommand{\theequation}{3.\arabic{equation}}
\setcounter{equation}{0}

\section{Stereographic projection} \label{Sect.Projection}
The surface of the sphere can be mapped by a stereographic projection
from the south pole ($\theta=\pi$) on the infinite plane tangent to 
the north pole ($\theta=0$). 
The complex coordinate in this plane is
\begin{equation}
z = 2 R \tan\left( \frac{\theta}{2} \right) {\rm e}^{{\rm i}\varphi} .
\end{equation}
The surface element (\ref{element}) transforms as
\begin{equation} \label{sigmastereo}
{\rm d}\sigma = \frac{{\rm d}^2 r}{\left( 1+\frac{r^2}{4 R^2} \right)^2}
\end{equation}
and the angular distance $\tau_{12}$ between points 1 and 2 on the sphere,
see Eq. (\ref{angular}), is given by
\begin{equation}
2 R \sin\left( \frac{\tau_{12}}{2} \right) =
\frac{\vert z_1-z_2 \vert}{\left( 1 + \frac{z_1\bar{z}_1}{4 R^2}\right)^{1/2}
\left( 1 + \frac{z_2\bar{z}_2}{4 R^2}\right)^{1/2}} .
\end{equation}

The application of the stereographic projection to the grand partition
function (\ref{Xi}) leads to the standard interaction Boltzmann factors
of the 2D Coulomb potential multiplied by one-body Boltzmann weights 
which depend on the specie type.
For the particle with positive charge $+e$ at position ${\bf r}$, one has
\begin{eqnarray}
w_+(r) & = & \frac{1}{\left( 1+\frac{r^2}{4 R^2} \right)^2}
\frac{1}{\left( 1+\frac{r^2}{4 R^2} \right)^{\Gamma(Q+NZ-1)/2}}
\left( 1+\frac{r^2}{4 R^2} \right)^{\Gamma NZ/2} \nonumber \\
& = & \frac{1}{\left( 1+\frac{r^2}{4 R^2} \right)^{2+\Gamma(Q-1)/2}} , \label{wp}
\end{eqnarray}
where the rhs of the first line involves successively the contribution
from the surface element transformation (\ref{sigmastereo}), the
effect of the remaining $Q+NZ-1$ positive $+e$ charges and the effect of
of $N$ negative $-Ze$ charges.
Similarly, for the particle with negative charge $-Ze$, considering the effect 
of $Q+NZ$ positive $+e$ charges and the effect of the remaining $N-1$ 
negative $-Ze$ charges, one obtains
\begin{eqnarray}
w_-(r) & = & \frac{1}{\left( 1+\frac{r^2}{4 R^2} \right)^2}
\frac{1}{\left( 1+\frac{r^2}{4 R^2} \right)^{\Gamma (N-1) Z^2/2}}
\left( 1+\frac{r^2}{4 R^2} \right)^{\Gamma (Q+NZ)Z/2} \nonumber \\
& = & \frac{1}{\left( 1+\frac{r^2}{4 R^2} \right)^{2-\Gamma Z(Z+Q)/2}}  \label{wm}.
\end{eqnarray}
 
The grand partition function (\ref{Xi}) is rewritten in the planar format as
\begin{subequations} \label{Xistereo}
\begin{eqnarray} 
\Xi(\lambda_+,\lambda_-,R) & = & \sum_{N=0}^{\infty} 
\frac{\lambda_+^{Q+NZ}}{(Q+NZ)!} \frac{\lambda_-^N}{N!}
\int \prod_{i=1}^{Q+NZ} {\rm d}r_i^+\, w_+(r_i^+)
\prod_{i=1}^N {\rm d}r_i^-\, w_-(r_i^-) \nonumber \\ & & \times W_{Q+NZ,Z} , 
\end{eqnarray}
where the interaction two-body Boltzmann factor
\begin{equation}
W_{Q+NZ,Z} = \frac{\prod_{(i<j)=1}^{Q+NZ} \vert z_i^+-z_j^+ \vert^{\Gamma}
\prod_{(i<j)=1}^N \vert z_i^--z_j^- \vert^{\Gamma Z^2}}{
\prod_{i=1}^{Q+NZ} \prod_{j=1}^N \vert z_i^+-z_j^- \vert^{\Gamma}} .
\end{equation}
\end{subequations}

As concerns the planar number densities of species, in the expansion
(\ref{Xistereo}) we introduce for each term with $Q+NZ$ charges $+e$ and 
$N$ charges $-Ze$ the microscopic species number densities
\begin{equation}
\hat{n}_+({\bf r}) = \sum_i \delta({\bf r}-{\bf r}_i^+) , \qquad
\hat{n}_-({\bf r}) = \sum_i \delta({\bf r}-{\bf r}_i^-) .
\end{equation} 
Within the grand canonical formalism, the mean number densities of species
are defined as the averages 
\begin{equation}
n_{\pm}({\bf r}) = \langle \hat{n}_{\pm}({\bf r}) \rangle .
\end{equation}
At the same time, with regard to the Jacobian (\ref{sigmastereo}) of 
the stereographic projection, the constant densities on the sphere 
$n_{\pm}^{(s)}$ are transformed to the position-dependent ones on the plane
$n_{\pm}({\bf r})$ with the polar symmetry,
\begin{equation}
n_{\pm}(r) = \frac{n_{\pm}^{(s)}}{\left( 1+\frac{r^2}{4 R^2} \right)^2} .
\end{equation}
The number of particles is invariant with respect to the stereographic
projection, as follows directly from the equality
\begin{equation}
\int_0^{\infty} {\rm d}r\, 2\pi r n_{\pm}(r) = N_{\pm}^{(s)} .
\end{equation}
Note that the planar number densities at the $r=0$ origin coincide with 
the ones on the sphere,
\begin{equation} \label{origin}
n_{\pm}(0) = n_{\pm}^{(s)} .
\end{equation}

\renewcommand{\theequation}{4.\arabic{equation}}
\setcounter{equation}{0}

\section{Linear response} \label{Sect.Response}
It is useful to use a potential representation of the one-body
Boltzmann weights (\ref{wp}) and (\ref{wm}):
\begin{equation}
w_{\pm}(r) = {\rm e}^{-\beta V_{\pm}(r)} .
\end{equation}
For large $R$, the one-body potentials $V_{\pm}$ exhibit the leading behaviors
of the form
\begin{subequations}
\begin{eqnarray}
\beta V_+(r) & \sim & \left[ \left( 1 - \frac{\Gamma}{4} \right) 
+ \frac{\Gamma Q}{4} \right] \frac{r^2}{2 R^2} , \\
\beta V_-(r) & \sim & \left[ \left( 1 - \frac{\Gamma Z^2}{4} \right) 
+ \frac{\Gamma Z Q}{4} \right] \frac{r^2}{2 R^2} .
\end{eqnarray}
\end{subequations}

In the strict $R\to\infty$ planar limit, there is no external potential and 
the species number densities $n_{\pm}$ are uniform in space.
Taking into account that for a finite $r$ the potentials $V_{\pm}(r)$ 
are small in the large-$R$ limit, we intend to make a linear-response 
perturbation of our inhomogeneous system around the homogeneous planar one.
Let us denote by $\langle\cdots\rangle_0$ the thermal average over the
planar system with no external one-body potential in order to distinguish
it from the thermal average $\langle\cdots\rangle$ with one-body potentials 
$V_{\pm}$ included.
Writing
\begin{equation}
\sum_i V_+(r_i^+) + \sum_i V_-(r_i^-) = \sum_{q'=\pm} \int {\rm d}^2r'\,
\hat{n}_{q'}({\bf r}') V_{q'}({\bf r}') , 
\end{equation}
the mean species densities are expressible as
\begin{equation}
\langle \hat{n}_q({\bf r}) \rangle =
\frac{\langle \hat{n}_q({\bf r}) 
{\rm e}^{-\beta\sum_{q'=\pm}\int {\rm d}^2r'\,\hat{n}_{q'}({\bf r}') V_{q'}({\bf r}')}\rangle_0}{
\langle
{\rm e}^{-\beta\sum_{q'=\pm}\int {\rm d}^2r'\,\hat{n}_{q'}({\bf r}') V_{q'}({\bf r}')}\rangle_0} ,
\qquad q=\pm .
\end{equation}
Expanding this relation linearly in $V_{\pm}$, we arrive at
\begin{equation}
n_q({\bf r}) - n_q = - \sum_{q'=\pm} \int {\rm d}^2r'\,
\langle \hat{n}_{q}({\bf r}) \hat{n}_{q'}({\bf r}') \rangle_0^{\rm T}
\beta V_{q'}({\bf r}') ,
\end{equation}
where
\begin{equation}
\langle \hat{n}_{q}({\bf r}) \hat{n}_{q'}({\bf r}') \rangle_0^{\rm T}
= \langle \hat{n}_{q}({\bf r}) \hat{n}_{q'}({\bf r}') \rangle_0 - n_q n_{q'}
\end{equation}
is the truncated bulk two-body density of species $q$ at point ${\bf r}$ and 
species $q'$ at point ${\bf r}'$.
Setting ${\bf r}$ at the origin $0$, using the relation (\ref{origin})
and denoting by
\begin{equation}
I_{qq'} = \int {\rm d}^2 r\, r^2 \langle \hat{n}_{q}(0) \hat{n}_{q'}({\bf r}) 
\rangle_0^{\rm T}
\end{equation}
the second moment of the bulk two-body densities of species, we find 
in the large-$R$ limit that
\begin{eqnarray} \label{dif}
n_q^{(s)} - n _q & \sim & - I_{q+} \left[ \left( 1 - \frac{\Gamma}{4} \right) 
+ \frac{\Gamma Q}{4} \right] \frac{1}{2 R^2} ,  \nonumber \\ & & 
- I_{q-} \left[ \left( 1 - \frac{\Gamma Z^2}{4} \right) 
+ \frac{\Gamma Z Q}{4} \right] \frac{1}{2 R^2} .
\end{eqnarray} 

The explicit form of the second moments $I_{qq'}$ for the planar asymmetric TCP
was derived by using the renormalized Mayer expansion 
\cite{Deutsch74,Friedman62} in Ref. \cite{Samaj01}.
The asymmetric TCP was defined in that work as a mixture of charges $+1$
and $-1/Q$.
To match our notation with Eqs. (4.31) of Ref. \cite{Samaj01}, we have 
to identify $1\to +$ and $2\to -$, to substitute $\beta$ by 
$\beta e^2 = \Gamma$ and $Q$ by $1/Z$, with the result
\begin{subequations}
\begin{eqnarray}
I_{++} & = & - \frac{2 (3\Gamma Z^2-8)(\Gamma Z^2-6)}{3\pi\Gamma
(\Gamma Z-4)^2(Z+1)^2} , \\
I_{--} & = & - \frac{2 (3\Gamma-8)(\Gamma-6)}{3\pi\Gamma
(\Gamma Z-4)^2(Z+1)^2} , \\
I_{+-} & = & \frac{2 \left[ 3\Gamma^2 Z^2-2\Gamma (6Z^2-Z+6)+48\right]}{
3\pi\Gamma (\Gamma Z-4)^2(Z+1)^2}  
\end{eqnarray}
\end{subequations}
and $I_{-+} = I_{+-}$.

Substituting these explicit formulas for the second moments into 
Eq. (\ref{dif}), we obtain
\begin{equation}
\left( n_+^{(s)}-n_+ \right) - Z \left( n_-^{(s)}-n_- \right)
= \frac{Q}{4\pi R^2} .
\end{equation}
This equality is consistent with the previous couple of equations
(\ref{rovnice}) for the number density deviations which is a check
of the formalism.
Taking either $(n_+^{(s)}-n_+)$ or $(n_-^{(s)}-n_-)$ and comparing 
Eqs. (\ref{rovnice}) and (\ref{dif}), we find in the large-$R$ limit that
\begin{equation}
R \frac{\partial}{\partial R} f(R) \sim 
- \frac{1}{3} + \frac{1}{2} \Gamma Q^2 
\end{equation}
is independent of $Z$, as was anticipated.
Consequently,
\begin{equation}
f(R) \sim \left( - \frac{1}{3} + \frac{1}{2} \Gamma Q^2 \right) \ln R .
\end{equation}
Finally, substituting $f(R)$ into (\ref{finite}) and considering 
the definition of the grand potential (\ref{grandpot}),
we end up with the finite-size expansion (\ref{univ}) with the coefficients 
$A_V = - 4\pi (\beta P)$, $A_S=0$ and $B$ given by (\ref{B4}).

\renewcommand{\theequation}{5.\arabic{equation}}
\setcounter{equation}{0}

\section{Conclusion} \label{Sect.Conclusion}
The aim of this work was to study the effect of charge non-neutrality of 
a 2D Coulomb system on the finite-size expansion of its grand potential, 
in particular on the prefactor to the logarithmic term.
The previous studies of the symmetric TCP of mobile $\pm e$ charges at 
the collapse point $\Gamma=\beta e^2=2$ \cite{Ferrero14} and of 
the OCP (jellium) of equivalent mobile $e$ charges in the fixed neutralizing
background at the coupling $\Gamma$ being an even integer \cite{Samaj17} were 
restricted to the disk geometry of the confining domain with 
the Euler number $\chi=1$.
The previously obtained results suggest that the prefactor to the logarithmic 
term as a whole does not depend on the charge composition of the Coulomb system.
The prefactor consists in two terms: The first one $\chi/6$ is equivalent 
to the universal prefactor of neutral systems while the second one 
$-\beta (Q e)^2/2$ is non-universal.

To understand the effect of charge composition and domain geometry
on the non-universal term, we studied the asymmetric TCP of pointlike 
charges $+e$ and $-Z e$ $(Z=1,2,\ldots)$ on the surface of a sphere with 
$\chi=2$.
We used the special symmetry properties of the sphere, the stereographic
projection of the sphere onto an infinite plane combined with linear response 
theory \cite{Jancovici00a} and specific results for the second moments of 
the species density correlation functions \cite{Samaj01}.
The final result is surprising: The non-universal term is again
equal to $-\beta (Q e)^2/2$.
This fact indicates that this term does not depend neither on the composition 
of the Coulomb system (the valency parameter $Z$) nor on the domain topology
(the Euler number $\chi$).

There exist other relatively simple models, e.g., the symmetric TCP
in a disk at arbitrary $\beta e^2<2$, for testing our surmise.

Another task is to propose a general argument explaining the common form of
the non-universal term in the prefactor to the logarithmic finite-size
term for all kinds of 2D Coulomb fluids in an arbitrarily shaped domain. 
For the sphere domain of radius $R$ studied in this paper, a phenomenological 
type of assumption might be based on the fact that due to the rotational 
invariance the excess charge $Q e$ is spread uniformly over $\Lambda$ after 
thermal averaging, with the mean surface charge density $Qe/(4\pi R^2)$. 
With the Coulomb potential (\ref{spherepot}), the interaction 
excess-charge energy is given by
\begin{equation}
E = - \frac{1}{2} \left( \frac{Q e}{4\pi R^2} \right)^2
\int_{\Lambda} {\rm d}^2r \int_{\Lambda} {\rm d}^2r' \ln \left[ 
\frac{2 R}{L} \sin\frac{\tau({\bf r},{\bf r}')}{2}\right] ,
\end{equation} 
where $\tau({\bf r},{\bf r}')$ is the angular distance between the points
${\bf r}$ and ${\bf r}'$.
The separation of the $\ln R$-term is obvious and we end up with 
the large-$R$ result
\begin{equation} \label{final}
\beta E \sim - \frac{1}{2} \beta (Q e)^2 \ln R .
\end{equation}
This is the expected non-universal excess-charge contribution to 
the $\ln R$ term in $\beta\Omega$.

If $\Lambda$ is the 2D disk (logarithmic interaction), the repulsion of 
the excess charges causes their accommodation at the disk boundary with 
the line charge density $Qe/(2\pi R)$.
The interaction excess-charge energy
\begin{eqnarray}
E & = & - \frac{1}{2} \left( \frac{Q e}{2\pi R} \right)^2
\int_0^{2\pi} R {\rm d}\varphi \int_0^{2\pi} R{\rm d}\varphi' \ln \left[ 
\frac{2 R}{L} \sin\left(\frac{\varphi-\varphi'}{2}\right)\right] \nonumber \\
& = & - \frac{1}{2} (Q e)^2 \ln R  + {\rm cst.}
\end{eqnarray} 
coincides in the $R\to\infty$ limit with the previous one
(\ref{final}) for the sphere. 

In three dimensions, let us consider a system of charged particles interacting 
via the $1/r$ Coulomb potential, constrained to the interior of the sphere 
$\Lambda$ of radius $R$ and with the total excess charge $Q e$.
Due to repulsion to the sphere surface boundary $\partial\Lambda$,
the excess charges produce the homogeneous surface charge density 
$Qe/(4\pi R^2)$.
The self-energy of the surface excess-charge distribution is given by
\begin{equation}
E = \frac{1}{2} \left( \frac{Q e}{4\pi R^2} \right)^2
\int_{\partial\Lambda} {\rm d}\sigma \int_{\partial\Lambda} {\rm d}\sigma' 
\frac{1}{\vert {\bf r}-{\bf r}'\vert} ,
\end{equation}
where ${\bf r}=R(\sin\theta\cos\varphi,\sin\theta\sin\varphi,\cos\theta)$
and ${\rm d}\sigma$ is the surface sphere element (\ref{element}).
Rescaling ${\bf r}$ and ${\bf r}'$ by $R$ implies that $\beta E\propto 1/R$, 
i.e. the effect of the (finite) excess charge in the system on 
the grand potential is negligible in the limit $R\to\infty$.

Finally, we recall that the asymmetric TCP of pointlike charges $+e$ and $-Ze$
is thermodynamically stable against collapse of oppositely charged pairs
if $\beta Z e^2<2$ and all derivations and proofs presented in this work were
restricted to this temperature region.
As soon as $\beta Z e^2\ge 2$, the Coulomb particle interactions have to be
regularized at short distances, e.g. by hard-core potentials.
We would like to mention that the system stays in its fluid (conducting)
phase also in the region $\beta Z e^2\ge 2$, up to the Kosterlitz-Thouless 
(KT) transition to an insulating phase where our analysis does not apply;
it was suggested in Ref. \cite{Samaj01} that in the limit of a small 
hard-core radius the KT temperature is given by $\beta_{\rm KT} Z e^2 = 4$.      
It is not clear how the short-distance regularization of the Coulomb 
potential affects our results for the prefactor to the $\ln R$ term 
in the fluid phase $2\le \beta Z e^2<4$.

\begin{acknowledgements}
The support received from Grant VEGA No. 2/0003/18 is acknowledged. 
\end{acknowledgements}

\end{document}